\documentclass[utf8]{FrontiersinVancouver} 
\usepackage{url,hyperref,lineno,microtype,subcaption}
\usepackage[onehalfspacing]{setspace}
\usepackage[normalem]{ulem}

\nolinenumbers

\definecolor{colorA}{rgb}{0, 0.5, 0.5}
\definecolor{colorB}{rgb}{0.4, 0, 0.9}
\definecolor{colorC}{rgb}{0.9, 0, 0.4}
\definecolor{color_green}{rgb}{0, 0.39, 0}

\def\keyFont{\fontsize{8}{11}\helveticabold }
\def\firstAuthorLast{Kuchkin {et~al.}} 
\def\Authors{Vladyslav~M.~Kuchkin\,$^{1,2,3,*}$, Nikolai~S.~Kiselev\,$^{2,*}$, Filipp~N.~Rybakov\,$^{4}$ Igor~S.~Lobanov\,$^{5}$, Stefan~Blügel\,$^{2}$, and Valery~M.~Uzdin\,$^{6}$}
\def\Address{$^{1}$Science Institute, University of Iceland, 107 Reykjavík, Iceland \\
$^{2}$Peter Gr\"unberg Institute and Institute for Advanced Simulation, Forschungszentrum J\"ulich and JARA, 52425 J\"ulich, Germany\\
$^{3}$Department of Physics, RWTH Aachen University, 52056 Aachen, Germany\\
$^{4}$Department of Physics and Astronomy, Uppsala University, Box-516, SE-75120 Uppsala, Sweden\\
$^{5}$Independent Scholar, St. Petersburg, Russia\\
$^{6}$Independent Scholar, St. Petersburg, Russia}
\def\corrAuthor{Corresponding Author 1}

\def\corrEmail{vkuchkin@hi.is}

\begin{document}
\onecolumn
\firstpage{1}

\title{Heliknoton in a film of cubic chiral magnet}

\author[\firstAuthorLast ]{\Authors} 
\address{} 
\correspondance{} 
\extraAuth{Corresponding Author 2 \\ 
n.kiselev@fz-juelich.de
}

\maketitle

\begin{abstract}

\section{}
Cubic chiral magnets exhibit a remarkable diversity of two-dimensional topological magnetic textures, including skyrmions. 
However, the experimental confirmation of topological states localized in all three spatial dimensions remains challenging. 
In this paper, we investigate a three-dimensional topological state called a heliknoton, which is a hopfion embedded into a helix or conic background. 
We explore the range of parameters at which the heliknoton can be stabilized under realistic conditions using micromagnetic modeling, harmonic transition state theory, and stochastic spin dynamics simulations. 
We present theoretical Lorentz TEM images of the heliknoton, which can be used for experimental comparison. 
Additionally, we discuss the stability of the heliknoton at finite temperatures and the mechanism of its collapse.
Our study offers a pathway for future experimental investigations of three-dimensional topological solitons in magnetic crystals.

\tiny
 \keyFont{ \section{Keywords:}  heliknoton, hopfion, topological magnetic soliton, chiral magnets, stochastic LLG dynamics, micromagnetic simulations,  minimum energy path calculations, lifetime calculations} 
\end{abstract}

\section{Introduction}
Cubic chiral magnets have attracted significant theoretical and experimental attention due to the vast diversity of experimentally observed topological magnetic solitons. 
These materials include various Si- and Ge-based alloys with B20-type crystal structures, such as Fe$_{1-x}$Co$_x$Si~\cite{Yu_10, Park_14}, FeGe~\cite{Yu_11, Kovacs_17, Du_18, Yu_18}, MnSi~\cite{Yu_15}, and others~\cite{Shibata_13}. 
The competition between the Heisenberg exchange interaction and the chiral Dzyaloshinskii-Moriya interaction~\cite{Dzyaloshinskii, Moriya} (DMI) stabilizes topological solitons in these materials. 
The most extensively studied types of magnetic solitons in these systems are magnetic skyrmions~\cite{Bogdanov_89, Tokura2020}. 
Skyrmions are vortex-like strings or tubes characterized by a topological index $Q=-1$. 
Experimental observations have revealed that clusters of skyrmion tubes can form complex three-dimensional superstructures, such as skyrmion braids~\cite{Zheng_21}. 
Additionally, the skyrmion antiparticle, antiskyrmions, with topological charge $Q=+1$, have been observed in thin films of FeGe~\cite{Zheng_22}. 
Skyrmion bags with arbitrary topological charge and skyrmions with chiral kinks are other solitons that have been reported~\cite{Rybakov_19, Foster_19, Kuchkin_20ii}. 
Recently, the experimental observation of skyrmion bags with positive topological charge and their current-induced motion has been reported by Tang et al.~\cite{Tang_21}.

Skyrmions, antiskyrmions, and skyrmion bags are two-dimensional (2D) topological solitons that are localized in the plane of the sample and confined by the free surfaces of the sample in the third dimension. 
The homotopy classification of these solitons is based on the continuous mapping between two spheres, $\mathbb{S}^2\!\rightarrow\!\mathbb{S}^2$. 
Where the pre-image is a sphere that is homeomorphic to the two-dimensional region where the skyrmion is localized, while the image is a sphere that corresponds to a parameter space of the magnetization unit field, given by $\mathbf{n}(\mathbf{r})=\mathbf{M}(\mathbf{r})/M_\mathrm{s}$.
For the magnetic texture in $xy$-plane the corresponding topological charge~\cite{DubrovinFomenkoNovikov_2}:
\begin{equation}
Q = \dfrac{1}{4\pi}\int (\mathbf{F}\cdot\hat{\mathbf{e}}_z) \,\mathrm{d}x\mathrm{d}y, 
\label{Q}
\end{equation}
where the vector field 
\begin{equation}
\mathbf{F} = 
\left( 
\begin{array}{c}
\mathbf{n}\cdot[\partial_y\mathbf{n}\times\partial_z\mathbf{n}]\\ 
\mathbf{n}\cdot[\partial_z\mathbf{n}\times\partial_x\mathbf{n}]\\ 
\mathbf{n}\cdot[\partial_x\mathbf{n}\times\partial_y\mathbf{n}]   
\end{array}
\right)
\end{equation}
is the curvature vector of the vector field $\mathbf{n}$ and is frame-invariant, meaning it does not depend on the choice of coordinate frame~\cite{Aminov_1969, Aminov}. 
For an unambiguous definition of $Q$, the direction of the $z$-axis should be chosen consistently with the magnetization $\mathbf{n}_0$ around the region of texture localization such that $\hat{\mathbf{e}}_z\cdot \mathbf{n}_0>0$, see details in Refs.~\cite{Rybakov_thesis,hopfionring}.

Three-dimensional (3D) topological magnetic solitons belong to a distinct class of solutions that are localized in all three spatial dimensions. The classification of 3D magnetic solitons is based on the mapping $\mathbb{S}^3\!\rightarrow\!\mathbb{S}^2$, which was originally proposed by Hopf~\cite{Hopf}. The Hopf index $H$, which is the 3D analog of the topological invariant $Q$, was derived by Whitehead~\cite{Whitehead47} and is given by:
\begin{equation}
H=-\frac{1}{16\pi^2}\int(\mathbf{F}\cdot\mathbf{V})\mathrm{d}x\mathrm{d}y\mathrm{d}z,
\label{Hopf_index}
\end{equation}
where $\mathbf{V}$ is a vector potential satisfying $\nabla\times\mathbf{V}=\mathbf{F}$.
For localized 3D magnetic textures without singularities, the Hopf index $H$ is an integer that has the following meaning: 
two morphologically distinct magnetic textures with identical $H$ can be continuously transformed into each other without the appearance of singularities, whereas continuous transformations between magnetic configurations with different $H$ are impossible. 
Statically stable structures with non-zero Hopf index are commonly referred to as Hopf solitons or hopfions. 
It should be noted that the topological charge $Q$ is zero for any hopfion cross-section. 
An example of a hopfion texture with $H=1$ is depicted in the first column of Fig.~\ref{Fig1}.

Theoretical studies of hopfions in various models of magnetic crystals began in the 1970s and continue to the present day~\cite{DzyaloshinskiiIvanov, Bogolubsky88, PhysRevLett.82.1554, Sutcliffe2017, PhysRevB.98.174437, Voinescu_20, Rybakov_22, Sallermann_23, Lobanov_23}. 
There is also some limited experimental evidence for direct observation of such magnetic textures~\cite{Kent2021, hopfionring}.
An intriguing type of hopfion has been predicted in cubic chiral magnets~\cite{Voinescu_20}, but has not been observed so far. 
This hopfion is embedded not in a collinear ferromagnetic state but in a helical spin spiral state. We refer to it as a \textit{heliknoton} following the terminology used in Refs.\cite{TaiSmalyukh2019, Voinescu_20}. 

The primary goal of this study is to investigate the stability of the heliknoton under realistic conditions, taking into account the demagnetizing field effects, and estimate its stability range in films of different thicknesses. 
Additionally, we explore how the contrast in theoretical Lorentz transmission electron microscopy (TEM) images changes with film thickness and suggest optimal conditions for experimental observations of the heliknoton. 
Furthermore, we estimate the stability of the heliknoton at finite temperatures using two complementary methods: harmonic transition state theory (HTST)\cite{bessarab_2012,bessarab_2013} and stochastic spin dynamics~\cite{Landau_Lifshitz}.

\section{Model}

To estimate the stability of heliknoton in the films of finite thickness, we use a micromagnetic model, which contains the Heisenberg exchange term, the DMI, the Zeeman interaction, and the self-energy of the demagnetizing field~\cite{Zheng_21, MicromagneticsRevisited}:
\begin{equation}
\mathcal{E}\!=\!\!\int\limits_{\mathbb{R}^3}\!\!\left[\! \mathcal{A} 
|\nabla \mathbf{n}|^2 
\!+\!\mathcal{D}\,\mathbf{n}\!\cdot\!(\nabla\!\!\times\!\mathbf{n})
-\! M_\mathrm{s}\,\mathbf{n}\!\cdot\!\mathbf{B}\!+\! \dfrac{|\nabla\!\!\times\!\mathbf{A}|^{2}}{2\mu_0} 
\!\right]\!d\mathbf{r}\, ,\label{E_micromagnet}
\end{equation}
where ${\mathbf{n}(\mathbf{r})}$ is the magnetization unit vector field,  $M_\mathrm{s}$ is the saturation magnetization,
$\mathcal{A}$ is the exchange stiffness constant, $\mathcal{D}$ is the constant of isotropic bulk DMI, and $\mu_0$ is the vacuum permeability.
The magnetic field $\mathbf{B}(\mathbf{r})$ in \eqref{E_micromagnet} represents the sum of a homogeneous external magnetic field $\mathbf{B}_\mathrm{ext}$ and the demagnetizing field produced by the sample itself,
$
\mathbf{B} = \mathbf{B}_\mathrm{ext} + \nabla\!\times\!\mathbf{A},
$
where ${\mathbf{A}(\mathbf{r})}$ is the magnetic vector potential due to the presence of the magnetization field $\mathbf{M}(\mathbf{r})=M_\mathrm{s}\mathbf{n}$.
It should be noted that while the unit vector $\mathbf{n}$ is defined only within the volume of the sample, the vector potential $\mathbf{A}(\mathbf{r})$ is defined in the entire space of $\mathbb{R}^{3}$. 
As a result, the integration for the first three terms in Eq. \eqref{E_micromagnet} must be carried out only within the sample volume. 
For the present analysis, we will consider an infinite plate lying in the $xy$-plane and assume that the external magnetic field is parallel to the $z$-axis.

The static equilibrium heliknoton was obtained by numerically minimizing \eqref{E_micromagnet} with respect to the pair of fields $\mathbf{n}$ and $\mathbf{A}$ using a nonlinear conjugate gradient method (for details, see Ref.~\cite{Zheng_21}).
The simulated domain was discretized on a regular mesh with $256\times256\times181$ nodes along the $x$, $y$, and $z$ axes, respectively. 
To simulate an extended film, periodic boundary conditions were applied in the $xy$-plane. 
The results presented below for the micromagnetic calculations were obtained using GPU-accelerated software Excalibur~\cite{Excalibur}.
Additionally, the results were verified with the publicly available software MuMax3~\cite{Mumax}. 
For definiteness, we used the material parameters for FeGe~\cite{Zheng_18}:
$\mathcal{A}=4.75$~pJm$^{-1}$, $\mathcal{D}=0.853$~mJm$^{-2}$, and $M_\mathrm{s}=384$~kAm$^{-1}$. 
To ensure that our results are generally applicable, we utilize reduced units for the external magnetic fields and distances throughout the paper. 
Specifically, we employ units concerning the cone saturation field $B_\mathrm{D}=\mathcal{D}^2/2M_\mathrm{s}\mathcal{A}$ and the equilibrium period of the spin spiral at the ground state $L_\mathrm{D}=4\pi\mathcal{A}/\mathcal{D}$ (70nm for FeGe). 
As has been shown earlier~\cite{Zheng_18}, when accounting for demagnetizing fields, the critical field for cone saturation changes to $B_\mathrm{c}=B_\mathrm{D}+\mu_0M_\mathrm{s}$ (682 mT for FeGe), and $B_\mathrm{c}=B_\mathrm{D}$ (199 mT for FeGe) only if the demagnetizing field is neglected.

\section{Results}

\subsection{Initial guess}

\begin{figure*}[ht]
\centering
\includegraphics[width=18cm]{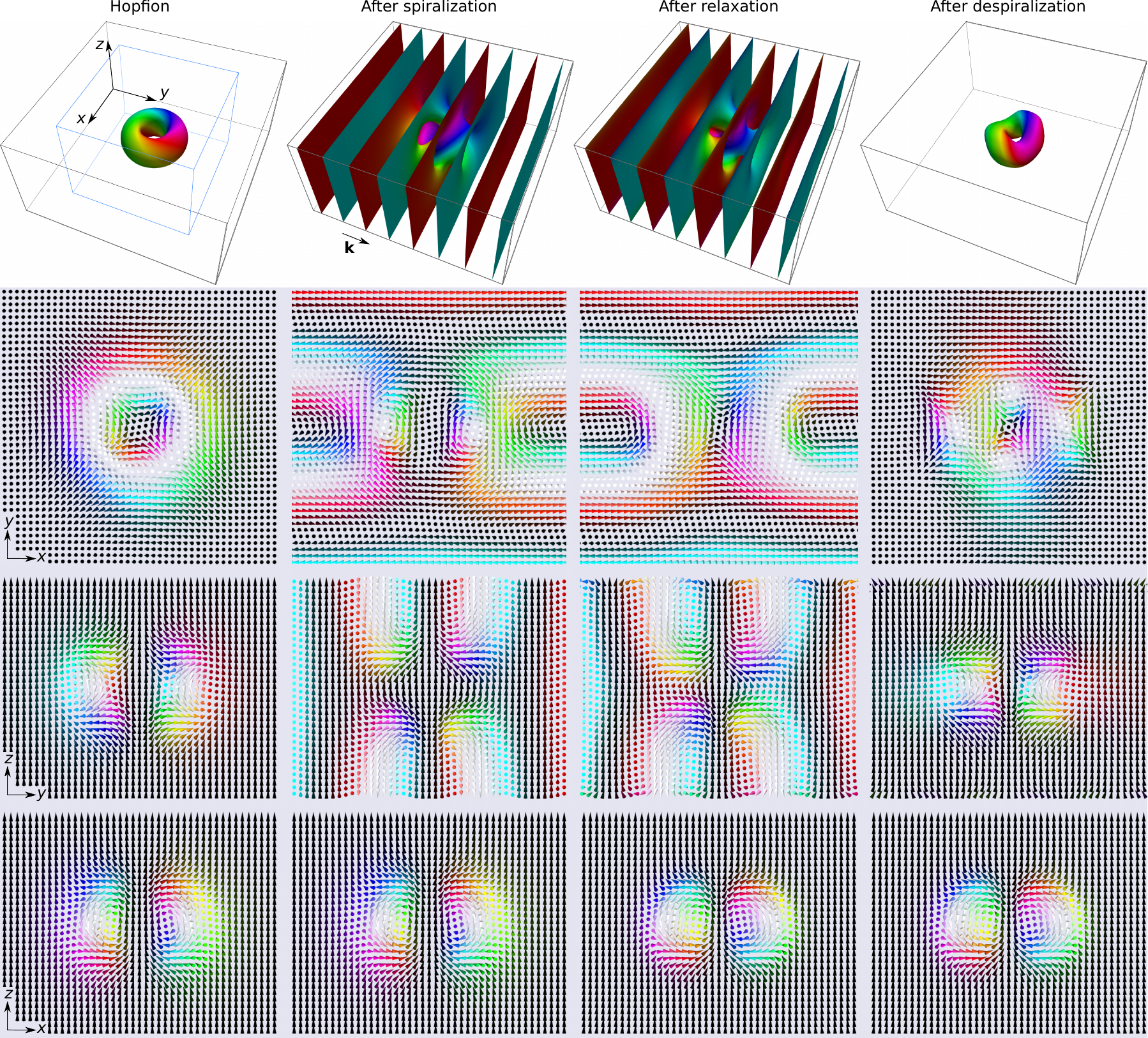}
\caption{\small 
The images illustrate the spin texture inside the simulated box of the size $4L_\mathrm{D}\times 4L_D$ in the $xy$-plane and thickness of $2L_\mathrm{D}$ along the $z$-axis.
The first column of images correspond to the hopfion anzatz \eqref{heliknoton_ansatz}.
The second column represents the hopfion ansatz after applying the spiralization \eqref{spiralization} with the $\mathbf{k}$-vector of the spiral parallel to the $y$-axis, $|\mathbf{k}|=2\pi/L_\mathrm{D}$.
The third column of images corresponds to the spin texture after the energy minimization assuming periodical boundary conditions in the $xy$-plane and free boundaries along the $z$-axis.
The fourth column illustrates the relaxed spin texture after applying despiralization.
The first row of images shows the isosurfaces $n_\mathrm{z}=0$.
The images depicted in the second, third, and fourth rows illustrate the magnetization field in the middle planes. For illustrative purposes, these images are bounded by the blue box of the size $0.64L_\mathrm{D}\times0.64L_\mathrm{D}\times2L_\mathrm{D}$ depicted in the top left image.}
\label{Fig1}
\end{figure*}

The initial configuration for heliknoton can be obtained in two steps.
First, we insert in the simulated domain the classical hedgehog-based ansatz~\cite{Hopf, Skyrme_1961} for a hopfion in the ferromagnetic background:
\begin{equation}
    \mathbf{n}=\dfrac{2\sin^{2}G}{r^2}\left(\begin{array}{c}x z\\-yz\\z^2\\\end{array}\right) + \dfrac{1}{r}\left(\begin{array}{c}y \sin2G\\x\sin2G\\ r\cos2G\\\end{array}\right),
    \label{heliknoton_ansatz}
\end{equation}
where function $G=G(r)$ with $r=\sqrt{x^2 + y^2 + z^2}$ describes the skyrmion-like profile: $G(r=0)=\pi$ and $G(r\rightarrow\infty)=0$. In our simulations we use $G(r)=2\arctan(\exp(-2 r)/r)$.
The first column in Fig.~\ref{Fig1} illustrates the spin texture obtained with the ansatz \eqref{heliknoton_ansatz}.
Note the opposite chirality of the spin texture in $x$ and $y$ directions.

In the second step, we implemented the spiralization procedure, which involves rotating all spins by an angle dependent on their coordinate along the wavevector $\mathbf{k}$ of the spiral. 
This transforms the ferromagnetic background into a helical one. 
Note that we applied this procedure to the entire volume, including the area containing the hopfion.
For the case $\mathbf{k}||\mathbf{e}_\mathrm{y}$ the spiralization procedure can be written as
\begin{equation}
    \mathbf{n}^\prime = \left(\begin{array}{ccc}
\cos \varphi & 0 & \sin \varphi\\
0 & 1 & 0\\
-\sin \varphi & 0 & \cos \varphi
\end{array}\right)\mathbf{n},\,\,\,\,\,\varphi=\dfrac{2\pi}{L_\mathrm{D}}y.
\label{spiralization}
\end{equation}
where $\mathbf{n}$ corresponds to the ansatz  \eqref{heliknoton_ansatz} and $\mathbf{n}^\prime$ is the ansatz for the hopfion in the helical background.
The second column in Fig.~\ref{Fig1} illustrates the spin texture obtained after applying the spiralization \eqref{spiralization} to the initial hopfion ansatz \eqref{heliknoton_ansatz}.
The spin texture after energy minimization is shown in the third column of Fig.~\ref{Fig1}.
The quality of our ansatz is seen from the comparison of the spin textures before and after energy minimization.
To better visualize the spin texture of the heliknoton, we utilize the despiralization procedure, which is illustrated in the last column of Fig.~\ref{Fig1}. 
This procedure serves as the inverse of the spiralization operation \eqref{spiralization}. 
After the energy minimization or capturing a snapshot of the magnetic texture, we apply the transformation \eqref{spiralization} with $\varphi\mapsto -\varphi$ to obtain the despiralized spin texture. 
By using this procedure, one can gain a better understanding of the heliknoton's spin texture.

\subsection{Demagnetizing field effect}

\begin{figure}[ht]
\centering
\includegraphics[width=8cm]{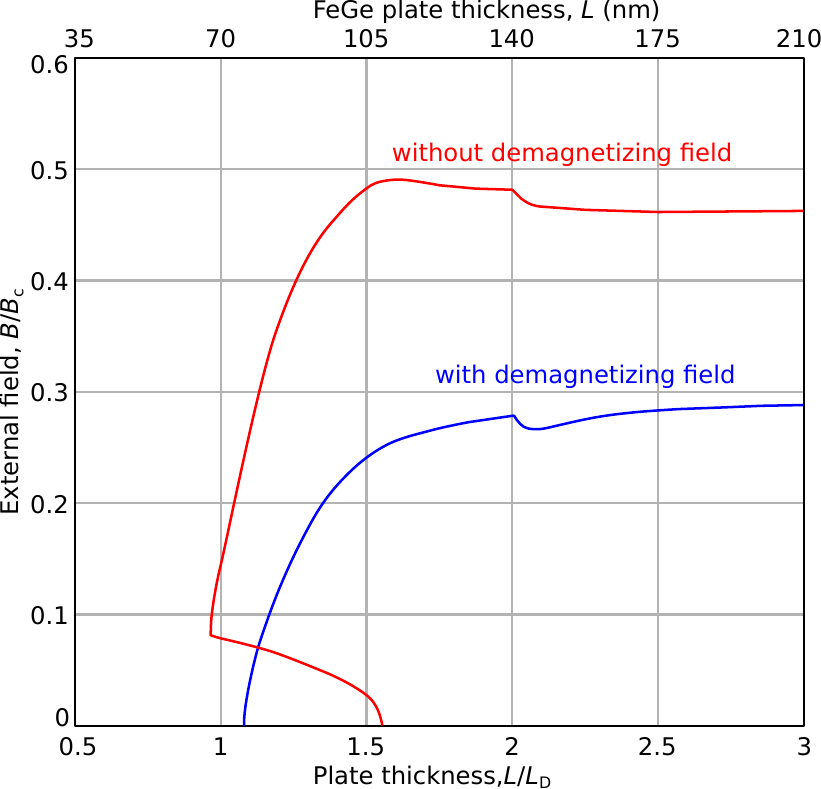}
\caption{\small 
The diagram of stability for heliknoton in extended film calculated with and without demagnetizing field. For consistency, the external fields for both cases are given in reduced units with respect to the saturation field of the cone phase, $B_\mathrm{c}=B_\mathrm{D}+\mu_0 M_\mathrm{s}$ and $B_\mathrm{c}=B_\mathrm{D}$, for the case with and without demagnetizing filed, respectively. For B20-type FeGe, $B_\mathrm{c}=B_\mathrm{D}+\mu_0 M_\mathrm{s}=0.682$ T and $L_\mathrm{D}=70$ nm, see the top axis.}
\label{Fig2}
\end{figure}

To estimate the demagnetizing field effect on the stability of heliknoton in a film of finite thickness, we performed systematic energy minimization at different external magnetic fields and film thickness starting with the ansatz \eqref{heliknoton_ansatz}-\eqref{spiralization}.
The results of these calculations performed with and without demagnetizing fields are presented in Fig.~\ref{Fig2}.
In these calculations, we keep the size of the simulated domain in $xy$-plane fixed, $4L_\mathrm{D}\times4L_\mathrm{D}$, and vary only the film thickness between $0.5L_\mathrm{D}$ and $3L_\mathrm{D}$.
Since the absolute values of the critical fields for the cases with and without demagnetizing fields are very different, we provide the critical fields in reduced units with respect to the saturation field of conical phase, $B_\mathrm{c}$.
This approach objectively estimates the range of heliknoton stability in both cases. 

As follows from the diagram in Fig.~\ref{Fig2}, taking into account the demagnetizing fields reduces the heliknoton stability range down to $\sim 0.3 B_\mathrm{c}$, which for the material parameters of FeGe corresponds to $\sim 200$ mT. This is a reasonably high range of fields accessible in most of the experimental setups.

At the film thickness below $L_\mathrm{D}$, the heliknoton does not fit the size of the film and becomes unstable in both cases, with and without demagnetizing fields.
With increasing thickness, the critical heliknoton fields tend to saturate in both cases.
Near the thicknesses commensurate to the period of helical modulations, \textit{e.g.}, $L/L_\mathrm{D}=2$, the critical fields show minor picks.
The most distinguishing is the behavior of the heliknoton in the thickness range between $1$ and $1.5L_\mathrm{D}$ without demagnetizing fields, where the solution remains stable only in the presence of the external field. 
At realistic conditions, with demagnetizing fields, however, it is not the case, and the heliknoton is stable in the whole range of $L>1.1L_\mathrm{D}$ even at zero magnetic fields.
The thickness of $\sim1.1L_\mathrm{D}$ can be thought of as the lower bound for experimental observations of heliknotons.

\subsection{Theoretical analysis of Lorentz TEM contrast}
According to the diagram shown in Figure~\ref{Fig2}, there is no upper bound limit for the sample thickness, and even in a bulk crystal, a heliknoton can exist. However, the thickness of the sample plays a crucial role in experimental observation. In particular, the sample's thickness is a significant factor for the applicability of Lorentz TEM.
In our previous studies~\cite{Zheng_21}, we estimated the upper limit for the thickness of the FeGe sample to be approximately 300 nm. Above this thickness, the sample becomes no longer transparent for electrons. 
In the case of thicker samples beyond 300 nm, one must use a TEM setup with an acceleration voltage above 300 kV, the standard voltage for the most modern TEM instruments.
Besides that, it is well known that for some magnetic textures, the Lorentz deflection forces can be completely canceled and thus give no contrast in non-tilted samples~\cite{Denneulin2021, Denneulin2021_2}.

\begin{figure*}[ht]
\centering
\includegraphics[width=18cm]{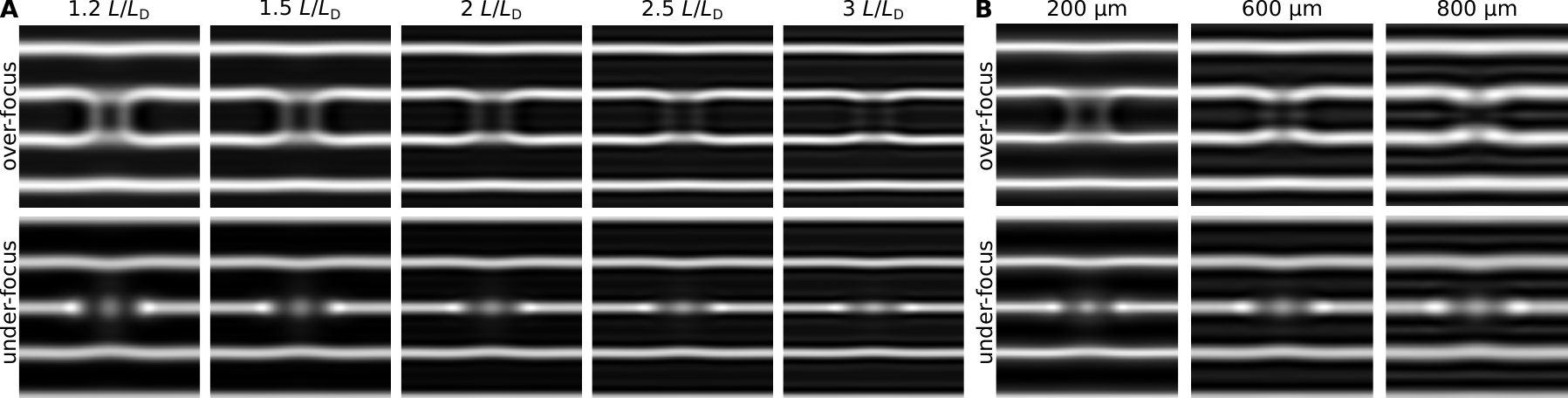}
\caption{\small 
(A) The theoretical Lorentz TEM images of heliknoton at zero external field in the plate of different thicknesses. The thicknesses in reduced units with respect to the period of helical modulations $L_\mathrm{D}$ are indicated on the top of each pair of over-focus and under-focus images. The defocus distance is 400 $\mu$m.
(B) The over-focus and under-focus Lorentz TEM images of heliknoton in the film of thickness $2L_\mathrm{D}$ calculated for different defocus distances.}
\label{Fig3}
\end{figure*}

Since the heliknoton is the texture localized in all three dimensions, it is natural to expect that the contrast it produces in Lorentz TEM is thickness dependent.
Figure~\ref{Fig3} shows the Lorentz TEM contrast provided by the heliknoton in films of different thicknesses, which we calculated assuming material parameters for FeGe. 
For these calculations, we used a well-established method based on the phase object approximation~\cite{MarcDe_Graef}. 
For details of implementation, see Ref.~\cite{Zheng_21}.

As seen from Fig.~\ref{Fig3}(A), above the thickness of $\ge2.5L_\mathrm{D}$ ($\sim 170$ nm for FeGe), the characteristic features in both over-focus and under-focus images of heliknoton are hardly seen.
The quality of the contrast can be improved by varying the defocus distance, as shown in Fig.~\ref{Fig3}(B).
However, in real experiments this is not always possible, mainly due to diffraction from the crystal structure, which usually also contributes to the Lorentz TEM contrast.  
Based on a comparison of the theoretical Lorentz TEM images and the stability diagram, the optimal thickness range for observing heliknotons in a TEM experiment is between $1.5L_\mathrm{D}$ and $2L_\mathrm{D}$ (around 100 nm and 140 nm, respectively, for FeGe).

\subsection{Minimum energy path calculations and HTST analysis}
To estimate the stability of heliknoton at finite temperature within HTST, we use an effective atomistic spin Hamiltonian defined on a simple cubic lattice:
\begin{align}
E =\!-J\!\sum_{\left\langle i,j\right\rangle}\!\mathbf{n}_{i}\cdot\mathbf{n}_{j} - \sum_{\left\langle i,j\right\rangle}\!\mathbf{D}_{ij}\cdot[\mathbf{n}_{i}\!\times\!\mathbf{n}_{j}] - \mu\sum_{i}\mathbf{B}_\mathrm{ext}\cdot\mathbf{n}_{i},
\label{Hamiltonian}
\end{align}
were $\mathbf{n}_i$ is the unit vector along the magnetic moment at the lattice site $i$, $J$ and $\mathbf{D}=D\hat{\mathbf{r}}_{ij}$ are the Heisenberg exchange constant and Dzyaloshinskii-Moriya (DM) vector, respectively, $\hat{\mathbf{r}}_{ij}$ is the unit vector along the segment connecting sites $i$ and $j$, $\mu$ is the magnitude of the magnetic moment at each site.
The symbol $\left\langle i,j\right\rangle$ denotes the summation over the nearest neighbor pairs over all lattice sites. 

Here we ignore the demagnetizing field effect and introduce characteristic parameters $L_\mathrm{D}=2\pi Ja/D$ and $B_\mathrm{D}=D^{2}/(J \mu)$ for the atomistic model \eqref{Hamiltonian}, where  $a$ is the lattice constant.

First, we apply HTST to estimate the lifetime of a particular magnetic state and the possible mechanisms of its collapse.
Within the HTST, the rate of transition between states $X$ and $Y$ at temperature $T$ is described by the Arrhenius law,
\begin{align}
 & k^{X\rightarrow Y}=\nu^{X\rightarrow Y}\exp{\left(-\dfrac{\Delta E^{X\rightarrow Y}}{k_\mathrm{B}T}\right)},\label{Arr}
\end{align}
where the energy barrier $\Delta E^{X\rightarrow Y}$ can be computed knowing the minimal energy path (MEP) connecting $X$ and $Y$ states on the energy surface as the energy difference between the highest point along the MEP -- the first-order saddle point ($SP$) on the energy surface of the system -- and the minimum at $X$. 
The pre-exponential factor $\nu^{X\rightarrow Y}$ incorporates dynamical 
$\nu_{dyn}$ and entropic $\nu_{ent}$ contributions to the transition rate
\[
\nu^{X\rightarrow Y} = {\frac{1}{2\pi}} \nu_{dyn}\nu_{ent}.
\]

Within the harmonic approximation, the energy of the system in the vicinity of a stationary state is approximated by a quadratic form that allows the prefactors to be determined explicitly~\cite{Lobanov21-CPC}.
The quadratic forms for the minimum $\mathbf{n}^X$ and the saddle point $\mathbf{n}^{SP}$ 
are determined by the corresponding Hessian matrices $\mathcal{H}^X$ and $\mathcal{H}^{SP}$, respectively.
The entropy of the states is expressed in terms of determinants of the matrices.
If the energy is an invariant of some transformation, e.g., the energy of the hopfion is preserved during translations, 
then some eigenvalues of the Hessians are zero, and the corresponding modes are called zero modes.
In the harmonic approximation, the contribution of zero-modes to the state $\mathbf{n}^X$ and saddle point $\mathbf{n}^{SP}$ which have $Z^X$ and $Z^{SP}$, respectively, the zero-modes  are estimated by their volumes $V^X$ and $V^{SP}$. 
The entropy prefactor is given by  
\[
\nu_{ent} = (2\pi k_BT)^{\frac{Z^{X}-Z^{SP}}2}\frac{V^{SP}}{V^X}
\sqrt{\frac{\det \mathcal{H}^{X}}{|\det \mathcal{H}^{SP}|}},
\]
where the Hessians are restricted to subspaces consisting of non-zero modes.

The dynamical prefactor is expressed in terms of
the negative eigenvalue $\zeta$ of the Hessian $\mathcal{H}^\mathrm{SP}$ with the corresponding eigenvector $e$,
\[
\nu_{dyn} = \sqrt{\frac{\mathbf{b}\cdot\mathcal{H}^\mathrm{SP}\mathbf{b}}{|\zeta|}},\quad 
\it b_i = \frac{\gamma\zeta}{\mu}\mathbf{n}_i^\mathrm{SP}\times e_i.
\]
The index $i$ numbers the lattice sites.

The heliknoton in bulk has at least two zero modes corresponding to translations in directions orthogonal 
to the helical axis.
We eliminate one of the zero modes by pinning the spins at the boundaries $x=\mathrm{const}$ and $y=\mathrm{const}$
in such a way that the direction of the magnetic moments coincides with one of the helix.
The pinned spins were not affected by the optimization procedures and were not included as degrees of freedom
to the Hessians.
On the surfaces $z=\mathrm{const}$ the boundary conditions are free,
which allows us to simulate the escape through the boundary.
Translation along the $z$-axis becomes a quasi-zero mode because of the effect of the boundary. 

The computation of MEP was performed in Cartesian coordinates with constraints on the magnetic moments length taking into account by introduction of Lagrange multipliers, which allows us to avoid singularities and the necessity to transform coordinates while changing the map in the atlas~\cite{LPU21}. 
The string method \cite{ERVE07} with stable tangent estimate \cite{HJ2000} was used for MEP calculation.

The most challenging part of the computation of the transition rate is the calculation of the determinants.
When a standard LU or QR decomposition
is used, the complexity of the problem is $\mathcal{O}(N^{3})$, where $N$ is the number of spins.
Taking into account the block band structure of the Hessian matrix for close-range interactions 
in the absence of the demagnetization field,
the complexity can be reduced to $\mathcal{O}(N^{7/3})$~\cite{Lobanov21-CPC}. 
Nevertheless, the problem remains time-consuming, especially for the domain of size  $5L_\mathrm{D}\times5L_\mathrm{D}\times2L_\mathrm{D}$ used in our calculations.

\begin{figure*}[ht]
\centering
\includegraphics[width=18cm]{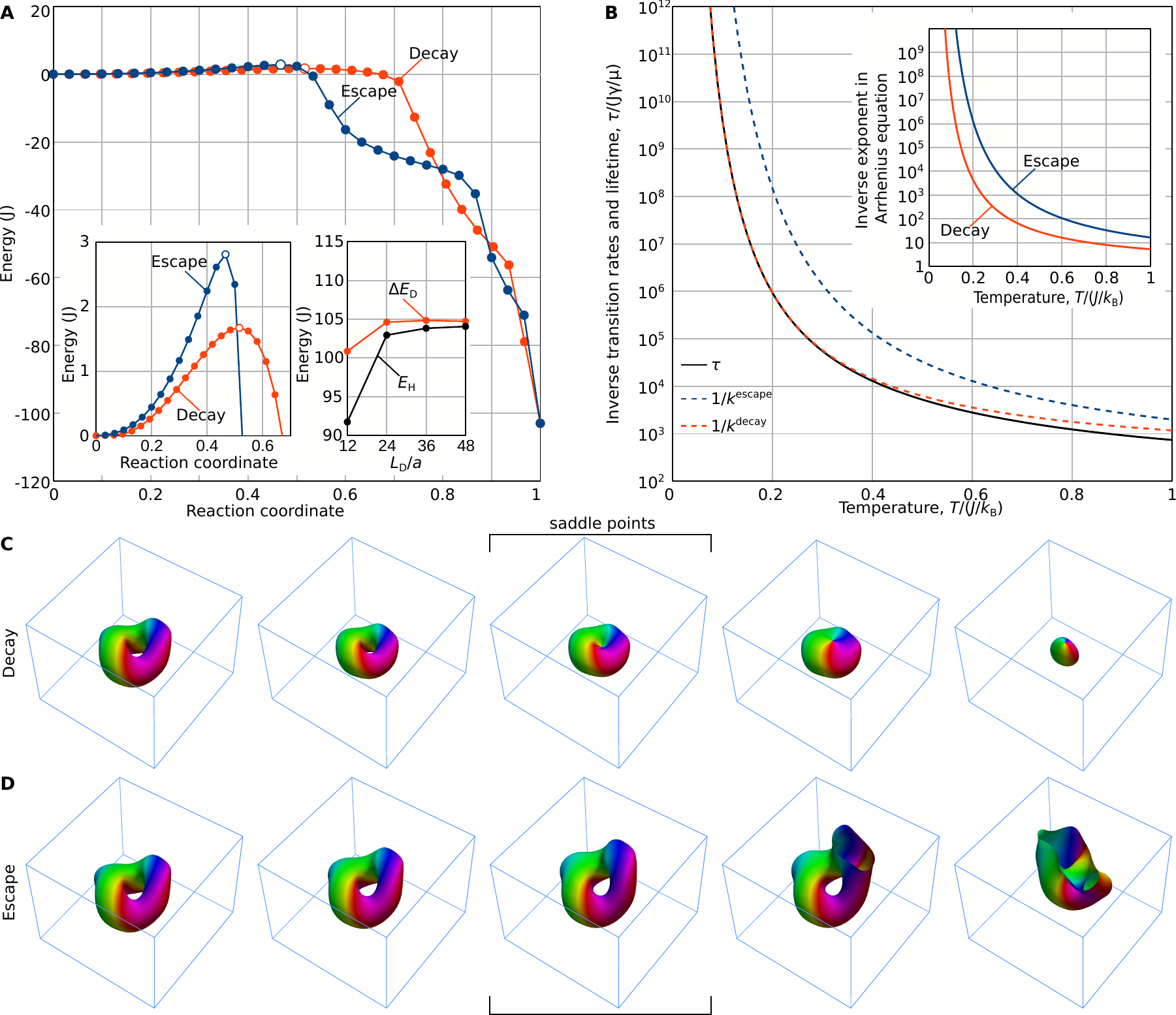}
\caption{\small 
(A) The minimum energy paths for two distinct mechanisms for heliknoton collapse in the film: the escape through the free edges of the plate (blue curve) and the decay inside the film via the nucleation of a pair of Bloch points (red curve).
The saddle points are indicated by hollow symbols.
The right inset illustrates the convergence of the energies of the heliknoton, $E_H$, and the energy barrier for the decay, $\Delta E_D$, to the micromagnetic limits with increasing mesh density, $L_\mathrm{D}/a$.
(B) The lifetime of heliknoton as a function of temperature.
The escape mechanism ( blue line) is less probable than the decay mechanism for all temperatures. 
Insets in (B) show the contributions of the exponent (activation barrier) in Arrhenius law to the lifetime.
(C) Snapshots of the system along the minimum energy path for heliknoton decay. The spin texture after despiralization is represented by the isosurfaces $m_\mathrm{z}=0$. The number above each image corresponds to the reaction coordinate, $r$. 
Similar to Fig.~\ref{Fig1}, the images show the volume confined by the blue box of the size
$2.5L_\mathrm{D}\times2.5L_\mathrm{D}\times2L_\mathrm{D}$ -- the quarter of the whole simulated domain of the size $5L_\mathrm{D}\times5L_\mathrm{D}\times2L_\mathrm{D}$.
(D) Snapshots of the system along the minimum energy path for heliknoton escape through the free surface. The field of view and notations are the same as in (C).
In (C) and (D), the images in the middle correspond to the saddle points of corresponding MEPs.
}
\label{Fig-HTST}
\end{figure*}

To find the optimal mesh density,
We computed MEPs calculated at different mesh densities of $L_\mathrm{D}$ equal to $12a$, $24a$, $36a$, and $48a$.
While the coarsest grid of $12a$ gives significantly different values of both the Heliknoton energy and the activation barrier,
the results for the three densest grids (with  $24a$, $36a$, and $48a$) are only slightly different (see right inset in Fig.~\ref{Fig-HTST}A).
Therefore, we used the discretization density $L_\mathrm{D}=24a$ for the transition rate computation.

The results of the HTST analysis are summarized in Figure \ref{Fig-HTST}. 
The MEPs presented in Figure \ref{Fig-HTST}A correspond to two distinct mechanisms for heliknoton collapse, which we refer to as decay and escape.
The decay mechanism is similar to the collapse of hopfions in a model of a frustrated magnet ~\cite{Sallermann_23,Lobanov_23}.
In this case, the heliknoton collapses via the emergence of a pair of magnetic singularities, namely Bloch points of opposite signs.
Representative images along the MEP corresponding to heliknoton decay are shown in Figure \ref{Fig-HTST}C.

As seen from the images in Figure \ref{Fig-HTST}D, when the heliknoton escapes through the boundary, magnetic singularities do not appear at any point along the MEP. 
Surprisingly, the escape mechanism for the heliknoton collapse is characterized by a higher energy barrier compared to heliknoton decay
for sufficiently dense lattices. 
Thereby, unlike the case of hopfions in the model of  frustrated magnets \cite{Rybakov_22}, the escape through the boundary for the heliknoton requires overcoming some activation barrier, which we attribute to the emergence of the chiral surface twist effect~\cite{Rybakov_15}.
The latter suggests that the heliknoton in a thin plate of a chiral magnet may be more promising for experimental observations due to its finite energy barrier for escape.

The total transition rate for a metastable state is the sum of transition rates for all mechanisms of the collapse.
The main contributions to the mean lifetime (inverse to the transition rate) of the heliknoton 
are the decay and escape mechanisms, with decay dominating:
\[
\tau = \frac1{k^{decay}+k^{escape}}\approx \frac1{k^{decay}}.
\]
The dependence of the heliknoton lifetime on temperature is shown in Figure \ref{Fig-HTST}B.
The inverse to the transition rate for the decay mechanism gives an estimate of the lifetime of the heliknoton in thick films
because the escape transition rate is inversely proportional to the film thickness. 

Overall, our HTST analysis sheds light on the decay mechanisms of the heliknoton and highlights potential avenues for its experimental observation, especially at low temperatures.
For instance, at $T = 0.455 J/k_\mathrm{B}$ (or $0.35T_\mathrm{c}$), the lifetime for heliknoton is $\tau\sim 10^6$ in dimensionless time units scaled
by $J\gamma \mu^{-1}$, see Fig.~\ref{Fig-HTST}B. 
For $J=112$~meV and $\mu=1030\mu_\mathrm{B}$, we obtain the heliknoton lifetime of $\sim 1000$ seconds, 
which is quite reasonable for experimental observation. 
Assuming every model spin $\mathbf{n}_i$ corresponds to a cluster of $7^3$ atoms,
where each atom has a magnetic moment of $3\mu_B$ and a lattice constant of $5$\AA,
the diameter of the observable heliknotons should be in the  order of $100$~nm.
Note that the temperature of $0.35T_\mathrm{c}$ for FeGe corresponds to $\sim 95$ K, which is accessible for the experiment performed at  liquid nitrogen temperature.

\subsection{Stochastic LLG dynamics}
The stochastic LLG equation can be written as
\begin{align}
 & \dfrac{\partial\mathbf{n}_{i}}{\partial t}=\mathbf{n}_i\times\left(\dfrac{1}{J}\frac{\partial E}{\partial \mathbf{n}_{i}}-\mathbf{B}_\mathrm{fluc}^{i}\right)+\alpha\mathbf{n}_{i}\times\frac{\partial\mathbf{n}_{i}}{\partial t},\label{LLG}
\end{align}
where $t$ is a dimensionless time scaled
by $J\gamma \mu^{-1}$, with $\gamma$ being the gyromagnetic ratio,
$\alpha$ is the Gilbert damping parameter,  $\mathbf{B}_\mathrm{fluc}^{i}$ is the fluctuating field representing uncorrelated Gaussian white noise with a prefactor proportional to the temperature, $T$.
For the numerical integration of Eq.~\eqref{LLG}, we use the semi-implicit method provided in Ref.~\cite{Mentink_10} implemented in the publicly available software Magnoom~\cite{Magnoom,Savchenko2022}.
For the chosen coupling parameters, we estimate the critical temperature, $T_\mathrm{c}\simeq 1.345 J/{k_\mathrm{B}}$ (see Ref.~\cite{Muller_2020}).
The simulations were performed with $\alpha=0.3$ and a fixed time step of $\Delta t=0.01$.
We perform ten independent simulations on a domain with $L_\mathrm{x}=L_\mathrm{y}=4L_\mathrm{D}$,  $L_\mathrm{z}=2L_\mathrm{D}$ and discretization density $L_\mathrm{D}/a=32$.
We use periodic boundary conditions in the $xy$-plane.
To increase the probability of heliknoton collapse in a reasonable time,
the simulations were performed at an elevated temperature of $T=1.0J/{k_\mathrm{B}}$ ($\sim0.74T_\mathrm{c}$).
The total simulation time of each run is $t=10^5$ ($10^7$ iterations).

\begin{figure*}[ht]
\centering
\includegraphics[width=17.6cm]{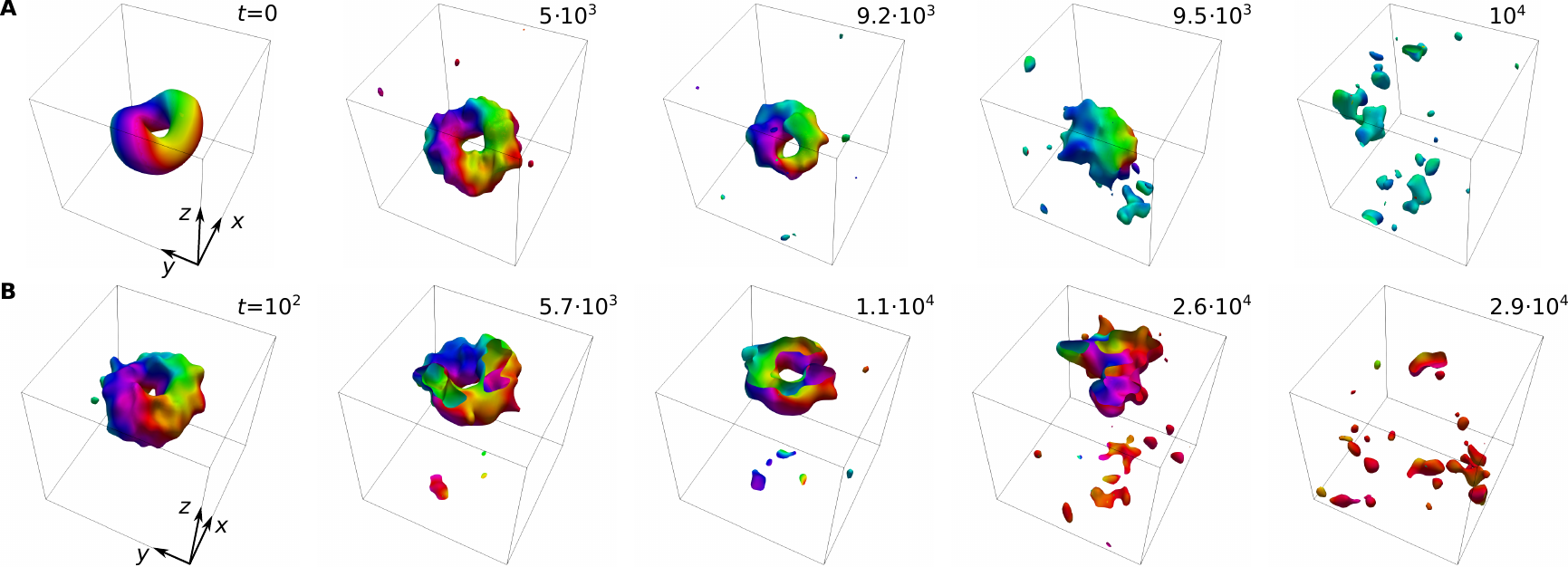}
\caption{\small 
Snapshots of the system in two different stochastic LLG simulations are shown. The set of images in row (A) corresponds to the heliknoton collapse through the nucleation of Bloch points, while images in row (B) correspond to its escaping through the free surface of the plate.
The shown box has size $2L_\mathrm{D}$ in each dimension.
}
\label{FigLLG}
\end{figure*}

To visualize the heliknoton isosurface $n_\mathrm{z}=0$ in the presence of strong thermal fluctuations, we first apply Fourier filtering as described in Ref.~\cite{Muller_2020}.
After that, we apply the despiralization procedure as introduced above, Fig.~\ref{FigLLG}.
Among the few independent runs of the LLG simulations, we chose the two most representative examples illustrated in Fig.~\ref{FigLLG}A, B and Supplementary Movie 1, 2.
In agreement with the results of HTST presented in the previous section, we observe two mechanisms of heliknoton collapse.
In particular, Fig.~\ref{FigLLG}A shows the snapshots of the systems during the heliknoton collapse via the decay, accompanied by the Bloch points nucleation.
Fig.~\ref{FigLLG}B illustrates the case when heliknoton collapses via escape through the top free surface of the plate.
These are the only two mechanisms observed in stochastic LLG simulations.
In the first case, the heliknoton shrinks until the tor hole completely disappears. At this moment, a pair of Bloch points of the opposite charge appears, and we see the transition from the heliknoton to the dipole string or toron~\cite{Muller_2020}.
The dipole string represents a non-stable state at these parameters and  collapses through shrinking.
In the second case, we observe the shifting of the heliknoton to the open surface as a result of its Brownian motion.
The escaping happens if the distance between the heliknoton and the surface is smaller than the critical one.
It is worth noting that despite the temporal fluctuations, the escape through the free surface represents a smooth transition.
Remarkably, the heliknoton remains stable near the surface of the plate for a noticeably long time. 
That might indicate the existence of a metastable state resembling surface modulations. 
Details of the investigation of such localized states will be provided elsewhere.

Among ten independent simulations, we observed the collapse of the heliknoton in the times range between $2\times10^4$ to $8\times10^4$ in reduced time units. Using these results, we estimated the lifetime of the heliknoton at $T=1.0J/k_\mathrm{B}$ to be approximately $\tau\sim10^4$. This result agrees with the findings of the HTST calculations presented in Fig.~\ref{Fig-HTST}B, which suggest a lifetime of $\tau\sim10^3$ for this temperature. 
The limited sampling (only ten independent runs) may account for the observed discrepancy. 
However, given the variation of the heliknoton's lifetime by ten orders of magnitude for different temperatures, we consider the agreement between the simulations and HTST calculations to be satisfactory.

\section{Conclusions}

In this study, we investigated the stability and decay mechanisms of the heliknoton in realistic conditions using direct energy minimization, stochastic LLG simulations, and HTST calculations. 
Our results showed that the heliknoton is stable in a wide range of magnetic fields and plate thicknesses, as demonstrated by the stability diagram calculated at zero temperature. 
However, our calculations indicated that for plate thicknesses above $2.5 L_\mathrm{D}$, the magnetic contrast in Lorentz TEM becomes too weak for reliable observations.

Based on our findings, we estimate the optimal plate thickness for heliknoton observations to be $2L_\mathrm{D}\pm0.5L_\mathrm{D}$ (140 $\pm$ 35 nm for FeGe parameters). 
Our analysis using the GNEB method revealed two main mechanisms of heliknoton decay: collapse via the formation of Bloch points and escape through the plate surface.
The LLG simulations at finite temperatures supported these observations and showed reasonable agreement with the estimation of heliknoton lifetime, at least at elevated temperatures. Moreover, the HTST calculations allowed us to estimate heliknoton lifetimes at different temperatures.
Overall, our results suggest that the heliknoton should be stable with a reasonably long lifetime at the standard measurement temperature of liquid nitrogen ($T = 95$ K) used in TEM experiments. 
We anticipate that these findings will stimulate further experimental observations of heliknotons in magnetic systems.

\section*{Conflict of Interest Statement}

The authors declare that the research was conducted in the absence of any commercial or financial relationships that could be construed as a potential conflict of interest.

 \section*{Author Contributions}

V.M.K., N.S.K. and F.N.R. performed micromagnetic and LLG simulations,  I.S.L. performed minimum energy path and lifetime calculations.  N.S.K. wrote the first draft of the manuscript. 
S.B. and V.M.U. participated in the interpretation of the results, review and editing, supervision. All authors contributed to the manuscript writing.

\section*{Funding}
V.M.K. acknowledges financial support from the Icelandic Research Fund (Grant No. 217750).
This project has received funding from the European Research Council under the European Union's Horizon 2020 Research and Innovation Programme (Grant No.~856538 - project ``3D MAGiC'').
N.S.K. and S.B. acknowledge financial support from the Deutsche Forschungsgemeinschaft through SPP 2137 ``Skyrmionics'', Grant No.\ KI 2078/1-1 and Grant No.\ BL 444/16, respectively. 
F.N.R. acknowledges support from the Swedish Research Council. 

\section*{Acknowledgments}

\section*{Supplemental Data}
Movies 1 and 2 illustrate two mechanisms of heliknoton collapse by decay and escape through the surface, respectively, observed in stochastic LLG simulations. 

\bibliographystyle{Frontiers-Harvard} 
\bibliography{test}

\section*{Figure captions}

\end{document}